\documentclass[pra,floatfix,twocolumn,superscriptaddress,showpacs,preprintnumbers,longbibliography,nofootinbib]{revtex4-1}
\usepackage{dcolumn}    
\usepackage{bm} 
\usepackage{graphicx}
\usepackage{amsmath}    
\usepackage{enumitem}
\usepackage{latexsym}
\usepackage{amsfonts}   
\usepackage{amssymb}
\usepackage{array}      
\usepackage{epsfig}
\usepackage{braket} 
\usepackage{bbold}
\usepackage{soul}
\usepackage[colorlinks=true,linkcolor=blue,urlcolor=blue,citecolor=blue,pdfusetitle]{hyperref}
\usepackage{hyperref}
\usepackage{amsthm}
\usepackage{soul}
\usepackage{mathtools}    
\usepackage{xcolor}

\newcommand{\ra}{\rangle}
\newcommand{\la}{\langle}

\newcommand{\eq}{Eq.~}
\newcommand{\eqs}{Eqs.~}
\newcommand{\fig}{Fig.~}

\DeclareRobustCommand\openzero{\leavevmode\hbox{0\kern-.55em0}}

\begin{document}

\begin{abstract}

Although the Schr{\"o}dinger and Heisenberg pictures are equivalent formulations of quantum mechanics, simulations performed choosing one over the other can greatly impact the computational resources required to solve a problem. Here we demonstrate that in Gaussian boson sampling, a central problem in quantum computing, a good choice of representation can shift the boundary between feasible and infeasible numerical simulability. To achieve this, we introduce a novel method for computing the probability distribution of boson sampling based on the time evolution of tensor networks in the Heisenberg picture. In addition, we overcome limitations of existing methods enabling simulations of realistic setups affected by non-uniform photon losses. Our results demonstrate the effectiveness of the method and its potential to advance quantum computing research.
\end{abstract}

\title{Simulating Gaussian Boson Sampling with Tensor Networks in the Heisenberg picture}

\author{Dario Cilluffo}
\affiliation{Institute of Theoretical Physics \& IQST, Ulm University, Albert-Einstein-Allee 11 89081, Ulm, Germany}

\author{Nicola Lorenzoni}
\affiliation{Institute of Theoretical Physics \& IQST, Ulm University, Albert-Einstein-Allee 11 89081, Ulm, Germany}

\author{Martin B. Plenio}
\affiliation{Institute of Theoretical Physics \& IQST, Ulm University, Albert-Einstein-Allee 11 89081, Ulm, Germany}

\maketitle

\section{Introduction}

Tensor networks \cite{Orus_2019, montangero2018introduction, Schollwoeck_2011} are a well-established methodology for providing a numerically efficient representation of quantum states, particularly effective in the domain of many-body physics. The complexity of the tensor network representation is fundamentally rooted in the bond dimension, which quantifies the quantum and classical correlations generated throughout the system during its evolution. Previous investigations, particularly within the context of a sudden quench applied to an Ising model, have proved that the simulation of wave function dynamics in the Schrödinger picture exhibits an exponential scaling in the bond dimension with respect to time \cite{Schuch_2008}. Conversely, the evolution of selected classes of observables in the Heisenberg picture manifests a linear scaling with time \cite{PhysRevLett.102.057202, clark2010exact, muth2011dynamical}. This observed contrast suggests the potential for enhanced computational efficiency in classical tensor network simulations by transitioning from the evolution of the wave function to the evolution of the desired observables.
In this paper, we analyse the impact of such scheme when simulating the dynamics of boson samplers.
\begin{figure}
\centering
\includegraphics[scale=0.3,angle=0]{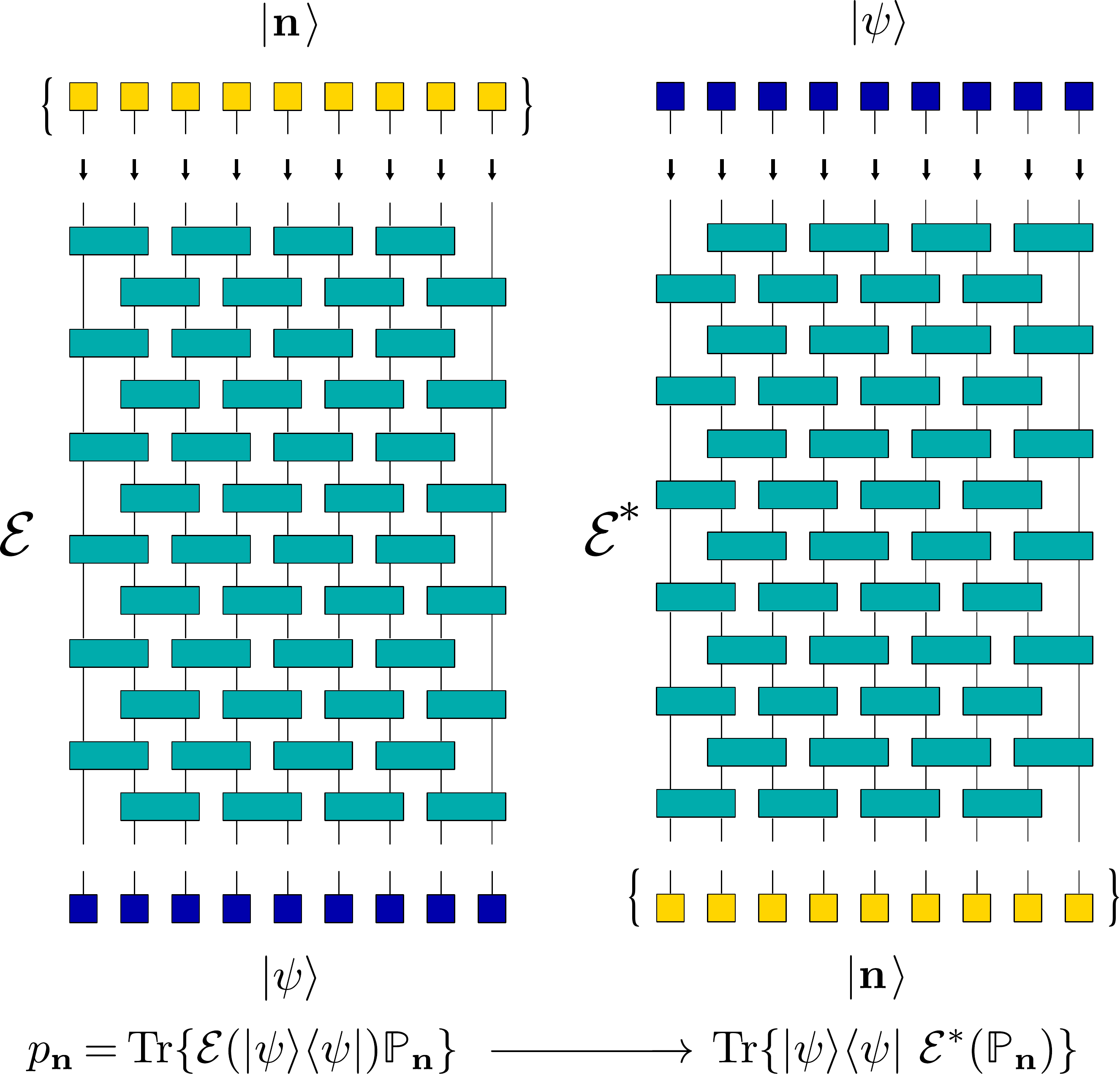}
\caption{In the Schrödinger picture (left) we evolve the initial state of the bosonic modes $\ket{\psi}$ through a network of elementary two-mode gates implementing the target transformation $\mathcal{E}$, thus we can project the evolved state over the manifold of the output states $\ket{\mathbf{n}}$. Conversely, in the Heisenberg picture (right) we compute the expectation value of each projector $\mathbb{P}_{\mathbf{n}}$ after the evolution through to the map $\mathcal{E}^*$.}
\label{fig1}
\end{figure} 

Boson sampling is a challenging problem in the broad field of quantum computing. On general grounds, it consists in generating samples of integer tuples based on a specific probability distribution that is known to be hard compute with classical computers \cite{AA} 
In its original formulation, the task is physically implemented by sampling the outcomes of a passive linear optical interferometer that is supplied with a Fock state containing a specified quantity of highly indistinguishable photons through input modes. This will be referred to as Fock Boson Sampling (FBS) throughout the text.
Yet, generating multimode highly indistinguishable input Fock states poses experimental difficulties, as it typically requires postselection from two-mode squeezed (Gaussian) states obtained through low-power down conversion, with a huge impact on scalability. This challenge has prompted the exploration of alternative input states that are readily accessible yet maintain the computational complexity of the simulation problem. One notable accomplishment in this area is Gaussian Boson Sampling (GBS) \cite{Ralph_2015,hamilton2017gaussian,Kruse}, which considers multi-mode squeezed states as input and does not require postselection, instead using the entire field generated by the sources as a computational resource. Although it is provably hard to classically simulate experimental boson sampling  when devices have none or only slight imperfections \cite{PhysRevLett.124.100502}, the extent to which current state-of the-art boson-samplers in the laboratory can achieve this regime remains an open question.

Nowadays, various numerical techniques have been employed in the simulations of boson sampling.
The main classical algorithms for simulating boson sampling \cite{Quesada_2020,PRXQuantum.3.010306,Bulmer,2017NatNeville} rely on conditioned probability chain rules with the aim to systematically generate high-quality output sequences element by element. 
However, this comes at the cost of calculating a limited set of analytical probabilities.
Phase space representation of linear optics has been successfully employed for verification of boson sampling experiments \cite{PhysRevA.105.012427}.
Tensor networks have also been used for simulating experimental quantum optics setups \cite{PhysRevA.97.062304}, in particular boson sampling experiments \cite{GarciaPatron}
, with a degree of approximation that tends to escalate with the increasing number of modes and indistinguishable photons. 
Notably, although tensor network architectures can entail, in general, a larger computational cost compared to directly computing the target probability distribution in the ideal case, their effectiveness becomes evident when photon correlations are significantly reduced, as observed in scenarios characterized by strong photon losses.
As a consequence, tensor networks have predominantly been applied to the development of well-scaling algorithms for simulating boson sampling in nontrivial yet markedly far-from-ideal scenarios reflective of the unavoidable limits imposed by current technology.
The original formulation of Fock and Gaussian boson sampling exclusively addresses the ideal scenario, where experimental imperfections are neglected. However, recognizing the significant impact of these imperfections in realistic setups, recent studies have been prompted to tackle these issues. Examples include addressing photon distinguishability \cite{PhysRevA.101.063840,shi2022effect} and handling photon losses. Photon losses, a major imperfection in modern devices, have thus far been analyzed using an approximate approach involving uniform dissipation, i.e.~every gate composing the interferometer is affected by the same imperfection.
This approach has allowed for an approximate assessment of the impact of dissipation, effectively transferring it to the input state of the interferometer, which could then be treated as lossless.
In instances of extreme dissipation, the quasi-thermality of the given input state leads to a reduction in the complexity of boson sampling, transforming its computational costs from exponential to polynomial \cite{Ralph_2015}. However, when photon losses are moderate, boson sampling remains a well-known exponentially complex problem \cite{PhysRevLett.124.100502}. In the Schr{\"o}dinger picture, in this scenario, it is well-stablished that multi-mode correlations in a boson sampler display exponential growth, akin to those observed following a sudden quench.
In this work, we employ a tensor network formalism to compute GBS probabilities in the Heisenberg picture and in presence of generic imperfections in the interferometer, specifically of moderate photon losses, when the problem is computationally challenging, and show how our method leads to an exponential advantage with respect the counterpart in the Schr{\"o}dinger picture. 
This work establishes the groundwork for future strides in refining sampling algorithms under analogous conditions, with the prospect of leveraging Markov-Chain MonteCarlo for potential enhancements.
While our method is specifically designed to address non-uniform losses, it also provides the capability to simulate a broader range of imperfections, encompassing non-Gaussian effects such as dephasing. These effects are often encountered in experimental scenarios and pose challenges that cannot be adequately addressed by existing methods.
We also note that it is also not solely restricted to the Gaussian boson sampling protocol, as in the Heisenberg picture we have the flexibility to explore various initial states, including correlated ones, with minor impact on the computational costs.

\section{Gaussian Boson Sampling in the Heisenberg picture}

We consider  a $M$-mode squeezed vacuum state $\ket{\psi} = \bigotimes_{j=1}^{M} \hat{S}_{r_j} \ket{0}_j$ inserted into an ideal $M$-mode linear interferometer, which can be represented (see section \ref{model}) as an $M \times M$ unitary mode-transformation matrix $\hat{\mathcal{U}}$. Given a a multi-mode $n$-photon Fock state $\ket{\mathbf{n}} = \ket{n_1,n_2,\ldots,n_M}$, with $n_k$ number of photons in the $k$-th mode, and the associated projector $\mathbb{P}_\mathbf{n} \coloneqq |\mathbf{n}\ra \la\mathbf{n} |$, the probability of measuring the state $\ket{\mathbf{n}}$ results 
\begin{align}
p_\mathbf{n} = {\rm Tr}\{ \mathcal{U}| \psi \rangle \langle \psi |\mathcal{U}^\dagger \mathbb{P}_\mathbf{n} \} =|\la \mathbf{n} \,|\, \hat{\mathcal{U}} \,| \, \psi \ra|^2 \equiv P_{\rm GBS}(\mathbf{n})\,.
\label{distr0}
\end{align}

Generating a sample of strings $\{\mathbf{n}\}$ distributed according to $P_{\rm GBS}$ identifies the Gaussian boson sampling problem \cite{Kruse}.
In such ideal case, computing the probability \eqref{distr0} for a $n$-photon outcome corresponds to the calculation of \textit{the Hafnian of a $n\times n$ complex matrix} determined by the evolved state $\hat{\mathcal{U}}\ket{\psi}$ (see Appendix \ref{app:haf}), which requires $\mathcal{O}(n^2 2^{n/2})$ arithmetic operations, rendering the related sampling problem intractable on classical computers.

Up to now, the main approach for simulating boson sampling with tensor networks consisted in evolving the multimode squeezed state $\ket{\psi}$ under the action of the unitary $\hat{\mathcal{U}}$, typically realized via an appropriate decomposition into two-body gates' networks (as depicted in \fig\ref{fig1}).
This approach leverages information about the evolved state to calculate the probability of a specific outcome $\mathbf{n}$ or by using it as a seed to generate marginal probabilities across a limited number of modes and exploiting chain rules to breed good output sequences \cite{GarciaPatron,complexity_TN}.
Here, in contrast, we compute probabilities in the Heisenberg picture by evolving the projectors $\mathbb{P}_\mathbf{n} \coloneqq |\mathbf{n}\ra \la\mathbf{n} |$, rather than the input state (see \fig\ref{fig1}):
\begin{align}
p_{\mathbf{n}} = {\rm Tr}\{  \mathcal{U} | \psi \rangle \langle \psi | \mathcal{U}^\dagger \mathbb{P}_\mathbf{n} \} \rightarrow {\rm Tr}\{ | \psi \rangle \langle \psi | ~ \mathcal{U}^\dagger \mathbb{P}_\mathbf{n} \mathcal{U} \}\, .
\end{align}

Noticeably, in the Heisenberg picture both FBS and GBS requires the evolution of multi-mode Fock states, and the different initial state only appears after the evolution has been performed.
An analytical form of $P_{\rm GBS}(\mathbf{n})$ is known only in the ideal case and for some restricted family of experimental imperfections. 
When generic imperfections are present the equation above generalizes to
\begin{align}
p_{\mathbf{n}} = {\rm Tr}\{ \mathcal{E}(| \psi \rangle \langle \psi |) \mathbb{P}_\mathbf{n} \} \rightarrow {\rm Tr}\{ | \psi \rangle \langle \psi | ~ \mathcal{E}^*(\mathbb{P}_\mathbf{n})  \}\, ,
\end{align}
where $\mathcal{E}$ is the dynamical map representing the action of the optical network on the input state.
In the tensor network framework, the difference between the two pictures is depicted in \fig \ref{fig1}: every component of the optical network is replaced by its time-reversed version, and the evolution is performed to each outcome instead of the multimode squeezed state.
Crucially, it's essential to note that evolving multimode Fock states and squeezed states is not generally computationally equivalent. This discrepancy results in a noteworthy computational advantage in the computation of Gaussian boson sampling (GBS) probabilities in the presence of unbalanced losses.

\section{Tensor network representation of a multiphoton interferometer} \label{model}

\subsection{Lossless multiphoton interferometer model}

We consider the tensor network representation of a $M$-mode GBS setup as introduced in \cite{PhysRevA.104.012415}.
The input state is represented by a tensor product of $M$ squeezed vacuum states
\begin{align}
\ket{\Psi} = \bigotimes_{i=0}^M \hat{S}(r_i) \ket{0}\,,
\label{sqst}
\end{align}
where $r_i$ is the squeezing parameter of the radiation in the $i$-th input mode and the squeezing operator reads
\begin{align}
\hat{S}(r_i) = {\rm exp}\{ r_i (\hat{a}_i^2-\hat{a}_i^{\dag 2}) \}\,
\end{align}
where $\{\hat{a}_i,\hat{a}_i^\dag\}$ are the ladder operators of the quantum harmonic oscillator representing the $i$th mode.
In the following we will assume for definiteness $r_i=r \, \forall i$.
The interferometer is implemented through the action of a series of 2-mode gates $\hat{G_{\theta, \varphi}}$, each constituted by a beamsplitter $\hat{U_{\theta, \varphi}}$ followed by a phase shift operation $\hat{P_{\phi}}$ on one mode as in
\begin{align}
&\hat{U}_{\theta, \varphi} = {\rm exp}\{ i \theta (\,\hat{a}_i \,\hat{a}_{i+1}^\dag e^{-i \varphi} + {\rm H.c.}) \}\,,
\label{eq:bs_gate}
\\&
\hat{P}_{\phi} = {\rm exp}\{ i \phi \,\hat{a}_i^\dag \,\hat{a}_{i} \}\,,
\\&
\hat{G}_{\theta, \varphi, \phi} = \hat{U}_{\theta \varphi} \cdot \hat{P}_{\phi} \,.
\end{align}

where $\theta, \varphi, \phi$ can be freely chosen for each gate.

With the aim of representing a realistic setup, we consider the interferometer as a sequence of $d$ layers, each constituted of $(M-1)/2$ 2-mode gates alternatively arranged (see \fig \ref{fig1}). 
In here, we set $d=M$, which does not harm generality thanks to the Clements protocol, as it allows to represent any unitary acting on $M$-modes as a circuit with depth $d=M$ through a polynomial-time algorithm \cite{Clements}.

\subsection{Losses and local dimension cutoff}

The standard way to describe a lossy optical mode $a$ is to couple it with an auxiliary mode $\eta$, which acts as an environment, through a beamsplitter and then operate a partial trace over the latter. More specifically, here we consider an auxiliary mode initially in the vacuum state $\ket{0}_\eta$, which yields a dynamical map $\mathcal{D}$ defined by a set of Kraus operators 
\begin{align}
\hat{K}_\mu = \,_\eta \la \mu | \hat{\mathcal{W}}_{a \eta} |0 \ra_\eta\,,
\end{align}
where we take $\hat{\mathcal{W}}_{a \eta} = \hat{U}_{\alpha,0}$ (see \eqref{eq:bs_gate}). 
This allows us to adjust the transmittivity $\gamma = {\rm sin}^2(\alpha)$ to control the magnitude of dissipation.
Without loss of generality, we model photon loss' effect at the level of each 2-mode gate by introducing dissipation on one of the two output modes.
As a consequence, the evolution of an observable defined over two modes $\la\hat{O}\ra\in \mathcal{H}^{\otimes 2}$ through a lossy gate reads 

\begin{align}
    \la\hat{O}'\ra 
    =
\rm{Tr}\Big\{ \sum_\mu \, \rho ~ \hat{\mathcal{U}}^\dag \hat{K}_\mu^\dag \hat{O} \, \hat{K}_\mu \hat{\mathcal{U}} \Big\}\,.
\label{gates}
\end{align}

As the Kraus map acting on $\hat{O}$ in the Heisenberg picture is $\mathcal{D}^*$, the resulting non-trace-preserving evolution will create, rather than dissipate, photons, typically increasing the required cutoff dimension of the mode's Hilbert spaces with respect the lossless scenario.
We can workaround this problem with a preliminary analysis of the probability distribution 
\eq \eqref{distr0} and with a simple statistical argument: in the Heisenberg picture each lossy gate is turned into an independent stochastic source of photons with rate $\gamma$. The 
probability of generating $x$ photons from $Q$ sources at rate $\gamma$ is well approximated 
by the binomial
\begin{align}
\pi_\gamma (x) = \binom{Q}{x} \gamma^x (1-\gamma)^{Q-x} \,,
\label{binomunif}
\end{align}
while the probability of generating $\nu$ couples of photons from M single-mode sources sharing 
the same squeezing amplitude $r$ reads \cite{Kruse}
\begin{align}
P_M^r (2\nu) = \binom{\nu + M/2 -1}{\nu} {\rm sech}^M(r){\rm tanh}^{2\nu}(r) \,.
\label{pdistr}
\end{align}
In the ideal case \eq \eqref{distr0} and \eq \eqref{pdistr} are related by
\begin{align}
P^{r}_{M} (\tilde{n}) 
=
\sum_{\mathbf{n}\in {\omega_{\tilde{n}}}}  p_\mathbf{n}\,,
\end{align}
where $\omega_{\tilde{n}}$ is the set of the outcomes $\mathbf{n}$ with fixed number of excitation $\tilde{n}$.
The same operation in a lossy setup yields
\begin{align}
P^{r}_{M\gamma} (\tilde{n}) =
\sum_{\mathbf{n}\in {\omega_{\tilde{n}}}}  p_\mathbf{n}^\gamma\,,
\end{align}
where $p^\gamma_\mathbf{n}$ is the unknown lossy GBS probability distribution.
In general for a fixed number of photons $\tilde{n}$, we expect to find $P^{r}_{M \gamma} (\tilde{n}) \neq P^{r}_M (\tilde{n})$ since the losses
create an unbalance between the probability to generate a $\tilde{n}$-photon component in input and the probability to observe an outcome configuration with the same number of photons. In particular losses will enhance the probability to observe outcomes with $\tilde{n}$ photons at the expenses of the probabilities of outcomes with $n>\tilde{n}$ photons.
Considering only the positive contribution, we easily find the upper bound
\begin{align}
P^{r}_{M \gamma}(\tilde{n}) - P^{r}_{M}(\tilde{n})
 \leq \Delta_\gamma (\tilde{n}) \coloneqq \sum_{x>0} \pi_\gamma(x) P^{r}_M (\tilde{n}+x) \,.
\label{dev}
\end{align}
For $\tilde{n}$ sufficiently large we expect this difference to vanish since $P^{r}_M(n)$ decays exponentially with $n$. On the other hand this implies that, in the presence of losses, the contribution of the states from subspaces with $n>\tilde{n}$ is exponentially suppressed.
In the Heisenberg picture this means that we can neglect the components of $\hat{O}'$ featuring more than $\tilde{n}$ photons with a maximum error of the order of $\Delta_\gamma (\tilde{n})$.
Thus we define the cutoff value ${n}_c$ for the local dimension of the oscillators as the value $n$ for which
\begin{align}
\Delta_\gamma(n={n}_c) < \epsilon\,,
\label{crit}
\end{align}
where $\epsilon \ll 1 $ is an arbitrary threshold.
In the experimental regime of interest
$\gamma$ is sufficiently small (significant losses can degrade the interference effects and make the problem more tractable for classical computers) and $d < \gamma^{-1}$, thus 
$\pi_\gamma(x)$ rapidly decays to zero and only few photons more than $\tilde{n}$ have to be considered for setting $n_c$.

\section{Bond dimension scaling}
In this section, we derive the scaling of the bond dimension required to accurately encode in a tensor network representation the evolution of the that permits exact representation of the 
Heisenberg operator. 
We can replace each projector $\mathbb{P}_\mathbf{n}$ with the corresponding quantum state $\ket{\mathbf{n}} = \prod_k {\hat{a}^{\dag \, n_k}_k}\ket{{\rm vac}}$. As a result, the problem can be transformed into investigating the time evolution of the creation operators.
Here we assume the total number of photons $n < M$  (this condition is typically required to ensure the computational hardness of boson sampling \cite{Kruse}). 
The propagation of a single photon in lossless a linear optical network featuring $M$ modes and at least $M$ layers
generates a delocalized single-photon state that reads
\begin{align}
\mathcal{U}(\hat{a}^\dag_k) \ket{\rm vac} = \sum_{i=0}^{M} u_{i k} \hat{a}^\dag_i \ket{\rm vac}\,,
\label{xstate}
\end{align}
where $\hat{a}_k^\dag = \bigotimes_{i=0}^{k-1} \mathbb{1}_{i} \otimes \hat{a}_k^\dag \otimes \bigotimes_{i=k+1}^M \mathbb{1}_{i}$.
With respect to any bipartition ($A,B$) of the support, an evolved single-photon state can be decomposed as
\begin{align}
\mathcal{U}(\hat{a}_k^\dag) = \sum_{j\in A} u_{k j} \hat{a}_j^\dag \otimes \mathbb{1}_B + \mathbb{1}_A \otimes \sum_{j\in B} u_{k j} \hat{a}_j^\dag \,,
\label{bd2}
\end{align}
that's equivalent to a W state.
Thus the operator $\mathcal{U} (\hat{a}^{\dag}_k) =  \hat{\mathcal{U}} \, \hat{a}^{\dag}_k  \,\hat{\mathcal{U}}^\dag$ can be represented as an ${\rm MPO}$ with maximum bond dimension $2$.
More in general, for $n$ photons in the same mode we have the multiphoton operator
\begin{align}
\mathcal{U}(\hat{a}_k^{\dag\,n}) = \sum_{k=0}^n \binom{n}{k} \bigg(\sum_{j\in A} u_{k j} \hat{a}_j^\dag \bigg)^{n-k} \!\!\!\!\!\! \otimes \bigg(\sum_{j\in B} u_{k j} \hat{a}_j^\dag \bigg)^k \,
\label{multiph}
\end{align}
that generates a state with at most $n+1$ independent components and is represented by an MPO with maximum bond dimension $D=n+1$.


In the case with $n$ input photons over a generic set of modes $s_m$ with size $m \leq n$ we have
\begin{align}
\mathcal{U} ( \prod_{k \in s_m} \hat{a}^{\dag\,n_k}_k) \ket{\rm vac}  = \prod_{k \in s_n} \mathcal{U} (\hat{a}^{\dag\,n_k}_k)  \ket{\rm vac}\,,
\label{ev_mpo}
\end{align}
where, in the last step, we have the product of $n$ MPOs each with maximum bond dimension $n_k+1$.
Thus for the maximum bond dimension of an MPS representing a $n$-photon state we find the bound
\begin{align}
D_{\rm max}^{\rm FBS} = \prod_{k \in s_m} (n_k + 1) \leq 2^n\,.
\label{bond_dim_post}
\end{align}
where the equality holds when $n=m$, i.e.~the upper bound is reached when all the input photons are injected through distinct ports.
This is somehow reminiscent of the fact that calculating the permanents of matrices containing repeated rows or columns may be comparatively simpler than calculating the permanents of non-repetitive matrices
\cite{aaronson2012generalizing}. Investigating this connection is beyond the scope of this paper, and it would be worthwhile to explore it in future research.
When $n \gtrsim M$ we observe a slight reduction of the bond dimension while remaining exponential in relation to the number of photons (see appendix A).

In the case of Gaussian input states the state \eqref{sqst} is injected through each mode and the number of photons is not fixed.
Therefore in this case we can span the whole space state until the bond dimension is saturated (see appendix A for details):
\begin{align}
D_{\rm max}^{\rm GBS} \sim  n^{M/2} \,.
\label{bond_dim_schr}
\end{align}
\begin{figure}
\includegraphics[scale=0.3,angle=0]{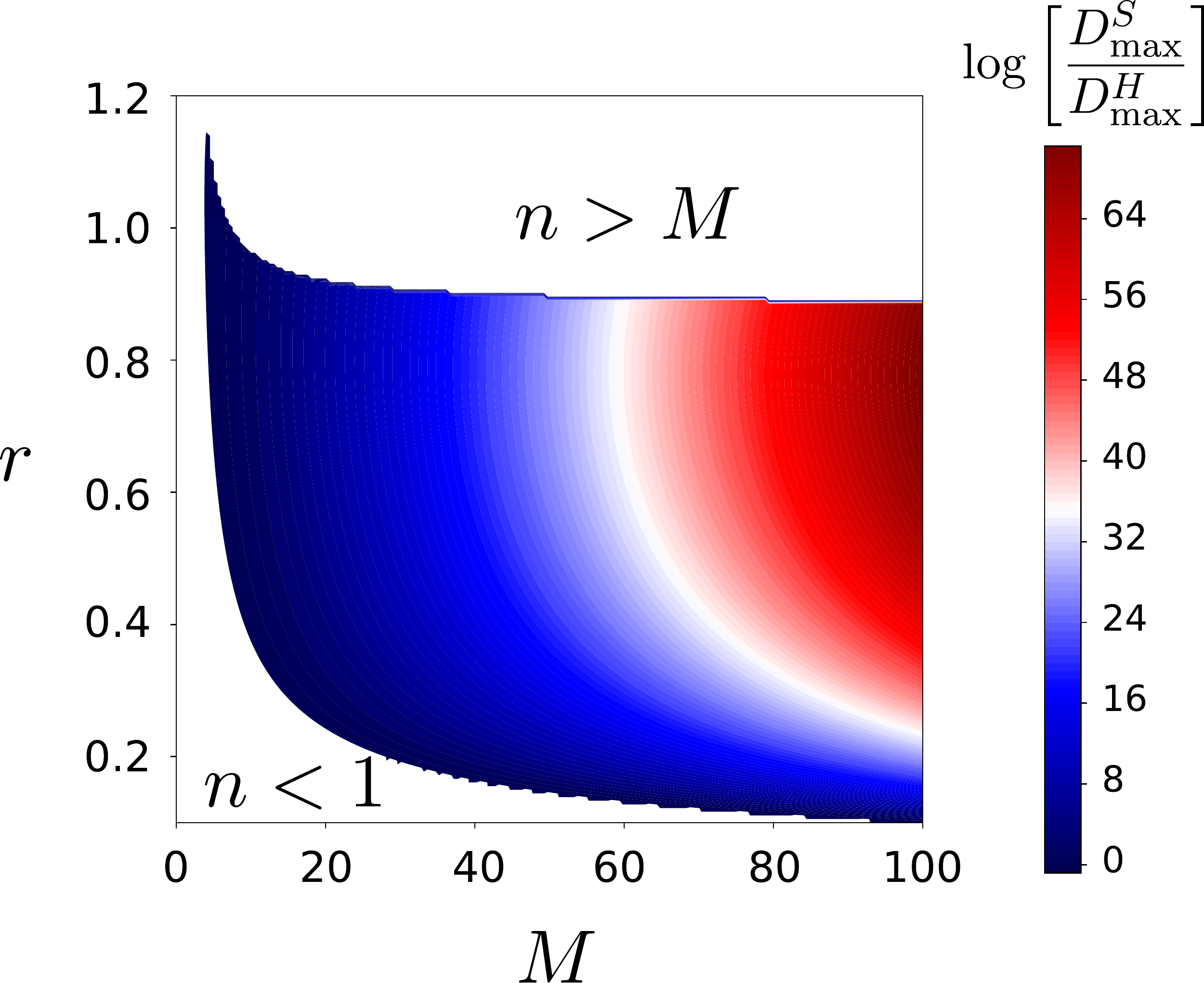}
\caption{\textbf{Maximum bond dimension scaling: Schrödinger vs Heisenberg in the ideal case.} We compare $D_{\rm max}^S$ and $D_{\rm max}^H$ (\eqs\eqref{bond_dim_schr} and \eqref{bond_dim_post}) as functions of the number of modes $M$ and the squeezing parameter $r$ in the regime of validity of the bond dimension scaling \eq\eqref{bond_dim_post} ($1<n={\rm mode}[P_M^r]<M$). 
}
\label{comp}
\end{figure}
The adoption of the Heisenberg picture reduces the task of computing a single bin $p_{\mathbf{n}}$ of the GBS probability distribution to a FBS taking as input $\mathbb{P}_{\mathbf{n}}$, followed by the projection over the squeezed input state, therefore in the Heisenberg picture $
D_{\rm max}^H\!=\! D_{\rm max}^{\rm FBS} $ (while in the Schrödinger picture $D_{\rm max}^S=D_{\rm max}^{\rm GBS}$).
Remarkably, the maximum bond dimension in the Heisenberg picture does not depend \textit{directly} on the number of modes $M$, but only on the number of photons $n$ in each probe output configuration $\mathbf{n}$. 
On the other hand, according to \eqref{pdistr}, the larger $M$, the more photons in the output configurations we must include in order to describe a significant part of the output distribution. 
A comparison between the scaling of the bond dimension in the Schrödinger and in the Heisenberg picture is reported in \fig\ref{comp}. For each couple of parameters $(r,M)$, the reference photon number is $n={\rm mode}[P_M^r] = 2(M/2 -1 ) {\rm sinh}^2(r)$. We restrict to the regime $1<{\rm mode}[P_M^r]<M$ for which \eqref{bond_dim_post} strictly holds. For $n \geq M$ photon bunching reduces the maximum bond dimension resulting in a better scaling (see Appendix A).
We can immediately see that, even for moderate squeezing, in the Schrödinger picture the bond dimension grows dramatically as $M$ increases. Conversely, the Heisenberg picture is advantageous in the near classically hard regime.


\section{Computational costs}
\label{sec:costs}
The computational costs on the proposed algorithm are upper-bounded by the complexity of the tensor network compression, namely $o(\chi_d^3\chi_o^2n_c^2)$ when considering losses. Here, $\chi_d$ and $\chi_o$ are the bond dimensions required for the tensor network representation of the evolved density matrix and gate operators, respectively, and $n_c$ denotes the considered number of photons. As dissipation terms correspond to gains in the Heisenberg picture, $n_c\geq n$ as a limited number of additional photons generated via these gains need to be considered. Regarding the bond dimension, we have shown that $\chi \leq 2^{n_c}$.
$\chi_o$, generally $\chi_o = n_c^2$, can be reduced to due to the U(1) symmetry, and particularly $\chi_o \to n_c$ when $n_c \gg 1$.
In the lossless case the complexity reduces to $o(\chi_s^3\chi_on^2)$, with $\chi_s \leq 2^n$ bond dimension required for the tensor network representation of the evolved state.


\section{Conclusions}

In this paper we have demonstrated that the Heisenberg picture may provide significant advantage in simulating the time evolution of bosonic many body systems. In the case of ideal Gaussian boson sampling, the evolution of squeezed vacuum states through passive linear optical networks yields a strongly correlated state whose numerical representation through MPS is in general very demanding in computational resources.
The scalings reported in Sec.~\ref{sec:costs} are notably worse than that of the best algorithms for computing Hafnians and permanents. This is not surprising since each evolved projector carries the information of a whole family of correlated Hafnians depending on the squeezed state we choose as input state and not only a single one. Our tensor network protocol allows to compute $p(\mathbf{n})$ in the presence of several types of experimental imperfections, in particular beyond the limit of uniform losses and Gaussian lossy channels, while an analytical closed form of the probability taking these effects into account is still unknown.
In such experimental conditions, this provides a groundbreaking chance to substantially improve classical sampling protocols in the same regime by employing the Markov-Chain MonteCarlo methods, which generally depend on directly computing a specific subset of analytical probabilities \cite{metropolis1953equation,geyer1992practical}. Importantly, the Metropolis-Hastings algorithm has demonstrated effectiveness in scenarios involving partial photon distinguishability \cite{PhysRevLett.120.220502}.
In addition, the Heisenberg picture evolution allows to draw the probability distribution of several types of boson sampling protocols through changes in the state $|\psi\ra$ (e.g.~for simulating Fock boson sampling) that are free of additional cost or switch to classical superpositions of projectors as observables (boson sampling with threshold detectors \cite{Quesada_2018}).
This versatility not only enhances the sampling capabilities but also opens up the potential for utilizing our algorithm in improving interferometer characterization protocols.
Even if computing each bin of the target probability requires the simulation of the action of the whole interferometer on each observable of interest, the scaling of the bond dimension (\fig \ref{comp}) makes this procedure rapidly advantageous with respect to the Schrödinger picture evolution in the nearly classically hard regime. 
Furthermore, the independent nature of individual tasks and the modularity of tensor networks enable different levels of parallelization, which can be efficiently exploited by utilizing GPU cluster computing. This is especially beneficial when dealing with a significant number of output configurations.

\section{Acknowledgements} 
This work was supported by the BMBF project PhoQuant (grant no. 13N16110) and the state
of Baden-Württemberg through bwHPC and the German Research Foundation (DFG) through grant no INST
40/575-1 FUGG (JUSTUS 2 cluster).
D.C. would like to thank T. Ilias and G. Di Meglio for fruitful discussions.

\appendix

\section{Hafnian}
\label{app:haf}
In the ideal case, the probability $P_{\rm GBS}(\mathbf{n})$ reads
\begin{align}
P_{\rm GBS}(\mathbf{n})= \frac{1}{\mathbf{n}! \sqrt{|\sigma_Q|}} {\rm Haf}(A_S)\,,
\label{distr}
\end{align}
where $\sigma_Q = \sigma + \mathbb{I}_{2 M}/2$, with $\sigma_{ij}= \la \{\hat{\zeta}_i,\hat{\zeta}_j^\dag \} \ra_\psi/2 -\la \hat{\zeta}_i \ra_\psi\la \hat{\zeta}_j \ra_\psi$ defined as the covariance matrix of the evolved state with $\{\hat{\zeta}\} = \{\hat{a}_1,\ldots,\hat{a}_M,\hat{a}^\dag_1,\ldots,\hat{a}^\dag_M\}$, and
\begin{align}
A_S= \Biggl( \begin{matrix} 0 & \mathbb{I}_M \, \\ 
\mathbb{I}_M \, & 0 \end{matrix} \Biggr)
 (\mathbb{I}_M - \sigma_Q^{-1})\, ,
\end{align}
The Hafnian of a generic $ 2 N \times 2 N$ 
square matrix $B$ is defined as \cite{caianiello1954quantum}
\begin{align}
{\rm Haf}(B) = \sum_{\mu \in {\rm PMP} } \prod_{j=1}^N {B}_{\mu(2j-1),\mu(2j)} \,,
\end{align}
with 
${\rm PMP}$ is the set of all the \textit{perfect matching permutations} of the string $ s = \{1,2,\ldots, 2N\}$, i.e.~all the permutations with the structure $\{a(1),b(1),a(2),b(2) \ldots a(N),b(N)\}$ with $a,b \in s$ such that $a(i) < b(i)\, \forall i$ and $a(i)<a(j)\,\forall i<j$ \cite{Callan}. 
Computing the Hafnian of a $N \times N$ complex matrix requires $\mathcal{O}(N^2 2^{N/2})$ arithmetic operations, rendering the related sampling problem intractable on classical computers.

\section{Bond dimension scaling for $n \gtrsim M$ and Schrödinger picture GBS}

In Section V, we made the assumption that $n < M$. However, when this condition is not met, the finiteness of the interferometer introduces boundary effects that require more sophisticated combinatorial considerations for accurate treatment. In this appendix, we provide an assessment of the upper limits on the bond dimension in this particular regime. As an additional result, the same approach will naturally yield an estimation of the scaling for Gaussian Boson Sampling (GBS) in the Schrödinger picture.
The MPO \eqref{ev_mpo} can be expressed as follows when considering a symmetric bipartition
\begin{align}
\mathcal{U} ( \prod_{k \in s_m} \hat{a}^{\dag\,n_k}_k)  = \sum_{k=0}^{N} \sum_{i} \lambda_i^{(k)} \hat{L}^i_k \otimes \hat{R}^i_{N-k} \,,
\label{st_test}
\end{align}
where the index $i$ runs over all the independent components with fixed number of photons $k$ and $N-k$ on the left and the right partition respectively, $\hat{L}_k$ ($\hat{R}_{N-k}$) is a product of $k$ ($N-k$) creation operators $\hat{a}^\dag_j$ and $M/2-k$ ($M/2-(N-k)$) identities $\mathbb{1}_j$ in all possible arrangements with $j \in [0,M/2-1]$($[M/2,M]$).
Note that when the number of photons in one partition is fixed, the number of photons in the other partition is automatically determined since the evolution preserves $N$.
The maximum bond dimension $D_{\rm max}$ is given by the number of coefficients $\lambda_i^{(k)}$ with multiplicity, i.e.~the sum over all the possible $k$ of the minimum number of independent components over the two partitions. Given $k$ input photons in one partition, each component corresponds to a subset of cardinality $k$ of a set of cardinality $M/2+k-1$ \cite{analytic_combinatorics}.
Thus a reasonable upward estimate is
\begin{align}
D_{\rm max} = \sum_{k=0}^N {\rm min}\left\{ \binom{m_L - 1 + k }{k} \,,\, \binom{m_R - 1 + (N-k) }{(N-k)}  \right\}\,,
\label{gen_estimate}
\end{align}
where $m_L$ and $m_R$ is the number of modes of the left and the right partition respectively and we consider the possibility of having an odd number of modes (i.e.~ $m_L$ and $m_R$ can differ by one).
Considering, for the sake of clarity, an even number of modes, so that $m_L=m_R = M/2$, we have
\begin{align}
D_{\rm max} = \begin{cases}
2 \sum_{k=0}^{N/2-1} \binom{\frac{M}{2}-1+k}{k} + \binom{\frac{M+N}{2}-1}{\frac{N}{2}} &\text{, $N$ even;}\\
2 \sum_{k=0}^{(N-1)/2} \binom{\frac{M}{2}-1+k}{k} &\text{, $N$ odd.}
\end{cases} 
\label{sc_mult}
\end{align}
Given that
\begin{align}
&\binom{\frac{M}{2}-1+k}{k} = \frac{M/2}{M/2+k} \binom{\frac{M}{2}+k}{k}\,,
\\
&\binom{\frac{M}{2}+k}{k} < \binom{N}{k} ~~ {\rm for}~ N \geq M\,,
\end{align}
and considering that
\begin{align}
2^N = \sum_{k=0}^{N} \binom{N}{k} =  2 \sum_{k=0}^{N/2-1} \binom{N}{k} + \binom{N}{N/2}\,,
\end{align}
we find $D_{\rm max}<2^N$.
The main hypothesis behind \eq\eqref{gen_estimate} is that the state is a Fock state with a fixed number of photons.
In the case of Schrödinger picture GBS, each component of the input state described in equation \eqref{sqst} possesses a distinct number of photons. Thus in this case the correspondence between the numbers of photons in the left and right partition does not hold. Consequently, assuming the local dimension cutoff $n_c$, in the Schrödinger picture each individual independent component corresponds to an $M/2$-tuple over the set $\{0,1,\ldots, n_c\}$. Therefore the total number of these components exactly matches the upper bound $n_c^{M/2} $\eqref{bond_dim_schr}.

\bibliography{biblio}

\end{document}